\documentclass[aps,prb,twocolumn,showpacs]{revtex4}

\usepackage{amsmath}
\usepackage{color}
\usepackage{amssymb}

\usepackage{graphicx}

\begin{document}

\title{Superconductivity in Uranium Compounds}

\author{V.P.Mineev}
\affiliation{Commissariat a l'Energie Atomique, INAC / SPSMS, 38054 Grenoble, France}

\begin{abstract}
On the basis of microscopic theory it is demonstrated how the
coupling between the electrons by means of magnetization fluctuations in ferromagnetic metal with orthorhombic symmetry gives rise  equal spin pairing superconducting state with general form of the order parameter dictated by symmetry. The   strong upturn of the upper critical field along 
$b$-direction above 5 Tesla in UCoGe is explained by the increase of pairing interaction caused by the suppression of the Curie temperature by magnetic field parallel to $b$-axis. It is proposed that similar phenomenon at much higher field must take place also for the field directed along the magnetically hardest a-direction.
\end{abstract}

\date{\today}
\maketitle
\section{Introduction}
Superconductivity arising in ferromagnet state of uranium compounds UGe$_2$, URhGe and UCoGe 
possessing extremely high upper critical field and the re-entrant superconductivity in large external magnetic field is obviously   related to triplet type superconductivity (for the most recent reviews see \cite{Aoki11,Aoki12,Aoki14}). 
These ferromagnets have orthorhombic structure with magnetic moments oriented along one of the crystallographic axis ($a$-axis in UGe$_2$, $ c$-axis in URhGe and UCoGe).
The general form of the order parameters in orthorhombic ferromagnet superconductors with triplet pairing allowed by symmetry  has been given in Ref.4. In neglect of interband pairing these superconducting states are reduced to the equal spin pairing states .\cite{Cham,Min04}

 It is quite natural to consider this type of superconductivity as magnetically mediated.  Here, however, there are two possibilities distinguished by the nature of ferromagnetic state.
Either it is due to spin-fluctuation exchange in itinerant electron ferromagnet, or due to electron  interaction with spin waves in system of ferromagnetically ordered localized magnetic moments.

Recently there was presented an argumentation\cite{Mineev2013,Chub} that ferromagnetism in uranium compounds is mostly supported by the moments located at uranium atoms. This case the most plausible pairing mechanism is due to interaction between the conduction electrons by means of spin waves in the system of localized moments. Such type model  applied to the 
ferromagnetic superconductor URhGe \cite{K.Hattori} successfully explanes the reentrant superconductivity phenomenon in this material.

 Even earlier there was developed a microscopic description of triplet superconductivity in ferromagnetic materials based on pairing interaction derived from phenomenological spectrum of fluctuations in the orthorhombic ferromagnet with strong magnetic anisotropy.\cite{Mineev2011}
 This approach allows to establish the interplay between pressure dependence of the Curie temperature and the critical temperature of superconducting transition. Also, there was demonstrated that 
depending of an external  field orientation parallel or perpendicular to the direction of spontaneous magnetization the effective  amplitude of pairing interaction proves to be decreasing or increasing function of magnetic field that allows to explain the drastic difference in magnitudes of upper critical field in these directions. 

Here we continue and develop further the approach of Ref.10. There will be presented more exact derivation of 
the magnetic susceptibility tensor and the amplitudes of  the pairing interaction. The  treatment of superconducting state will be done taking into account  both paramagnetic and orbital interaction of magnetic field with electron spins and charges.
We shall show how the coupling between the electrons by means of magnetization fluctuations in ferromagnetic metal with orthorhombic symmetry leads to formation of superconducting state with  general form of the order parameter dictated by symmetry established in Ref.5,6. 
The role of orthorhombic anisotropy originating of spin-orbital interaction in formation of superconducting state will be  demonstrated explicitly.

One of particular properties of UCoGe is the peculiar behavior of the upper critical field.\cite{Aoki2009}
Namely,  the upper critical field in $b$-direction reveals almost vertical upturn at fields around 5 tesla.
Strong up-curvature  is observed also  in $H_{c2}$  in $a$ direction but at lower temperatures  and higher magnetic fields.
In addition, at low temperature $T=0.09K$ which is much smaller than the zero-field superconducting temperature $T_{sc}=0.6K$ a steep angular dependence of $H_{c2}$ was reported when the field was slightly tilted from the $a$ axis toward the $c$ axis.\cite{Aoki2009}  This steep angular dependence of $H_{c2}$ cannot be explained at all with the anisotropic Ginzburg-Landau formula, even if the field dependence of effective mass $m^*$ is taken into account.

The description   of sharp angular dependence of the upper critical field was developed in frame of quite specific Ising type itinerant electron interaction approach.\cite{T.Hattory} Namely,  the paring interaction has been taken in form of wave vector-frequency dependent Fermi gas magnetic susceptibility. To stress the role of longitudinal fluctuations there was 
kept only $z$-component of susceptibility $\chi_{zz}({\bf q},\omega)$. The parallel to $c$ axis external field dependence of susceptibility has been chosen in nonanalytic form quite different from known mean-field  dependence. There was considered single band model with pairing only between spin-up electrons. The $H_{c2}$ angular dependence was obtained by solving the Eliashberg equations. The good correspondence of this description to the observations  is not astonished because it uses many 
 baseless assumptions specially chosen to get correspondence with experiment.

Here
 we describe qualitatively  the upper critical field temperature behavior as originating from  pairing interaction field dependence.  The   strong upturn of the upper critical field along 
$b$-direction above 5 Tesla in UCoGe is explained by the increase of pairing interaction caused by the suppression of the Curie temperature by magnetic field parallel to $b$-axis. Similar mechanism also stimulates the arising of the reentrant superconducting state in URhGe in fields along $b$-axis about 12 Tesla discovered about decade ago \cite{Levy}. It should be noted that the presented ideas about  the relationship between  ferromagnetic properties and   superconductivity  quite recently have been
  essentially supported  experimentally \cite{Hattory2014}.
 
 The paper is organized as follows. In Chapter II we introduce the microscopic Ginzburg-Landau-Gor'kov description of two-band  equal-spin pairing superconductivity
 arising in orthorhombic metal due to electron pairing by means of exchange of magnetic fluctuations. Then in Chapter III  we derive the susceptibility in orthorhombic ferromagnet that allows us to establish the field dependent pairing interaction. The structure of superconducting state is found in Chapter V. Finally, in Chapter VI we consider the field dependent effective critical temperature. The summary and discussion are contained in Conclusion.

\section{Ginzburg - Landau - Gor'kov equations}

We consider pairing originating from  the attraction
\begin{equation}
H_{int}=-\frac{1}{2}\mu_B^2I^2\int d^3{\bf r}d^3{\bf r}^\prime S_i({\bf r})\chi_{ij}({\bf r-{\bf r}^\prime})S_j({\bf r}^\prime)
\label{int}
\end{equation}
between  electrons with magnetic moments $\mu_B$ by means electron-magnon interaction in ferromagnetic media with orthorhombic symmetry.
Here, $
{\bf S}({\bf r})=\psi^\dagger_\alpha({\bf r})\mbox{\boldmath$\sigma$}_{\alpha\beta}\psi_\beta({\bf r})$ is the operator of electron spin density, $\chi_{ij}({\bf r}) $ is the media susceptiblity, $I\approx\frac{T_c}{nm^2}$ is an exchange constant, $T_c$ is the Curie temperature, n is the concentration of uranium atoms with zero-temperature magnetic moment $m$. 
The explicit form of pairing  interaction due to electron-magnon coupling  in weak coupling approximation is established  in the same manner as  it is done for electron-phonon coupling
(see fi book \cite{Kittel}).

The BCS pairing interaction due to electron-magnon coupling derived \cite{Samokhin} from Eq.(\ref{int}) is
\begin{equation}
H_{pairing}=\frac{1}{2}\sum_{{\bf k}{\bf k}^\prime}
V_{\alpha\beta\gamma\delta}({\bf k},{\bf k}^\prime)a^\dagger_\alpha({\bf k}) a^\dagger_\beta(-{\bf k}) a_\gamma(-{\bf k}^\prime) a_\delta({\bf k}^\prime),
\end{equation}
where
\begin{equation}
V_{\alpha\beta\gamma\delta}({\bf k},{\bf k}')=V_{ij}({\bf k},{\bf k}')
(i\sigma_i\sigma_y)_{\alpha\beta}(i\sigma_j\sigma_y)^\dagger_{\gamma\delta}
\end{equation}
and
\begin{equation}
V_{ij}({\bf k},{\bf k}')=-\mu_B^2I^2\left(\frac{1}{2}Tr\hat \chi^u({\bf k},{\bf k}')\delta_{ij}-\chi^u_{ij}({\bf k},{\bf k}')\right)
\label{3}
\end{equation}
is expressed   through  the odd part of media static susceptibility $\hat \chi^u({\bf k},{\bf k}')$
found in the next section.

The upper critical field is determined as the eigenvalue of the linear equation for the order parameter
\begin{eqnarray}
\Delta_{\alpha\beta}({\bf k},{\bf q})
=-T
\sum_{n}
\sum_{{\bf k}' }
V_{\beta\alpha\lambda\mu}({\bf k},{\bf k}')\nonumber\\
\times G_{\lambda\gamma}({\bf k}',\omega_n)
G_{\mu\delta}(-{\bf k}'+{\bf q},-\omega_n)\Delta_{\gamma\delta}({\bf k}',{\bf q}),
\end{eqnarray}
where the matrix of order parameter is
\begin{equation}
\hat \Delta=\left( \begin{array}{cc}\Delta^{{\uparrow}}& \Delta^{\uparrow\downarrow}\\ 
\Delta^{\uparrow\downarrow}& \Delta^{\downarrow}
\end{array}\right ),
\end{equation}
$G_{\lambda\gamma}({\bf k}',\omega_n)$
is the matrix of normal metal Green function. 
In  absence of external field  and when the magnetic field is parallel to the spontaneous magnetization it is  diagonal 
\begin{equation}
\hat G_n=\left( \begin{array}{cc}G^{{\uparrow}}& 0\\ 
0 & G^{\downarrow}
\label{matrix}
\end{array}\right ),
\end{equation}
where
\begin{equation}
G^{\uparrow,\downarrow}=\frac{1}{i\omega_n-\xi^{\uparrow,\downarrow}_{{\bf k}}
}.
\label{Gr}
\end{equation}

For the external field  ${\bf H}$ oriented either in the $(x,z)$  or in the $(y,z)$ plane , where the coordinates $x,y,z$ correspond to the $a,b,c$ crystallographic directions,
 it is natural to choose the spin quantization axis along the  direction of the total magnetic field 
 $(h+H_z)\hat z+H_x\hat x$ (or $(h+H_z)\hat z+H_y\hat y$) and to introduce the corresponding shift of chemical potential in the spin-up, spin-down bands. Here $h$ is the exchange field.  This case, also one needs  to write the interaction matrix in proper spin axis, as it is done in Ref.10. This makes the following treatment much more cumbersome. One can ignore this complicacy in assumption that  the exchange field is much larger than 
 an external field used in measurement of superconducting properties in uranium ferromagnetic superconductors \cite{footnote}. Under this assumption one can work with Eqs. (\ref{3}), (\ref{matrix}),
 (\ref{Gr}). Abandonment of this assumption leads to the appearance  of corrections  of the order of $\sim H^2_x/h^2$ or $\sim H^2_y/h^2$ to all obtained results.

 In view of large band splitting \cite{footnote} one may neglect the terms containing the products $G_1^\uparrow G_2^\downarrow +G_1^\downarrow G_2^\uparrow$.
Then the equations for the spin-up $ \Delta^\uparrow$ and the spin-down $\Delta^\downarrow$ order parameter components are 
\begin{widetext}
\begin{eqnarray}
\Delta^{\uparrow}({\bf k},{\bf q})
=-T
\sum_{n}
\sum_{{\bf k}' }
\left\{
V^{\uparrow\uparrow}({\bf k},{\bf k}')
G^\uparrow({\bf k}',\omega_n)
G^\uparrow(-{\bf k}'+{\bf q},-\omega_n)
\Delta^{\uparrow}({\bf k}',{\bf q})\right.\nonumber\\
+\left.
V^{\uparrow\downarrow}({\bf k},{\bf k}')G^\downarrow({\bf k}',\omega_n)
G^\downarrow(-{\bf k}'+{\bf q},-\omega_n)
\Delta^{\downarrow}({\bf k}',{\bf q})
\right\},~
\label{up}\\
\Delta^{\downarrow}({\bf k},{\bf q})
=-T
\sum_{n}
\sum_{{\bf k}' }
\left\{
V^{\downarrow\uparrow}({\bf k},{\bf k}')
G^\uparrow({\bf k}',\omega_n)
G^\uparrow(-{\bf k}'+{\bf q},-\omega_n)
\Delta^{\uparrow}({\bf k}',{\bf q})\right.\nonumber\\
+\left.
V^{\downarrow\downarrow}({\bf k},{\bf k}')
G^\downarrow({\bf k}',\omega_n)
G^\downarrow(-{\bf k}'+{\bf q},-\omega_n)
\Delta^{\downarrow}({\bf k}',{\bf q})
\right\}.
\label{down}
\end{eqnarray}
Here
\begin{eqnarray}
&V^{\uparrow\uparrow}({\bf k},{\bf k}')=V_{xx}+V_{yy}+iV_{xy}-iV_{yx}=-\mu_B^2I^2\chi_{zz}^u,
\label{11}\\
&V^{\downarrow\downarrow}({\bf k},{\bf k}')=V_{xx}+V_{yy}-iV_{xy}+iV_{yx}=-\mu_B^2I^2\chi_{zz}^u
,\label{112}\\
&V^{\uparrow\downarrow}({\bf k},{\bf k}')=-V_{xx}+V_{yy}+iV_{xy}+iV_{yx}=
-\mu_B^2I^2
(\chi^u_{xx}-\chi^u_{yy}-2i\chi^u_{xy}),
\label{113}\\
&V^{\downarrow\uparrow}({\bf k},{\bf k}')=-V_{xx}+V_{yy}-iV_{xy}-iV_{yx}
=
-\mu_B^2I^2
(\chi^u_{xx}-\chi^u_{yy}+2i\chi^u_{xy}).
\label{114}
\end{eqnarray}
\end{widetext}

One can see that the equations for $\Delta^{\uparrow}$ and for $\Delta^{\downarrow}$ are entangled. Hence, we deal with two-band superconducting state similar to $A_2$ state of superfluid $^3$He. This property is supported by the recent low temperature thermal conductivity measurements under magnetic field\cite{Howald}. It should be stressed that unlike to $^3$He where entanglement between spin-up and spin-down components  in absence of spin-orbit interaction  is absent that leads to  two subsequent phase transitions first to the spin-up and then to the spin-up-down superconducting states \cite{Mermin} these components in orthorhombic ferromagnet interact each other already in linear approximation that gives rise one common phase transition to the A$_2$ type state.
This is the particular property of superconducting state in an orthorhombic symmetry crystal with anisotropic susceptibility $\chi_{xx}\ne\chi_{yy}$. One band non-unitary superconducting state similar to $^3$He-$A_1$ is admissible  in an isotropic  metal with negligibly small spin-orbital coupling.

Let us now find the susceptibility.

\section{Magnetic susceptibility of orthorhombic ferromagnet}

The expression for static magnetic susceptibility is obtained following phenomenological approach of Ref.10.
The starting point is
the Landau free energy of orthorhombic ferromagnet 
in
magnetic field ${\bf H}({\bf r})$
\begin{equation}
{\cal F}=\int d V(F_M+F_\nabla),
\label{FE}
\end{equation}
where in 
\begin{eqnarray}
F_M=\alpha_{z}M_{z}¥^{2}+\alpha_{y}M_{y}^{2}+\alpha_{x}M_{x}¥^{2}~~~~~~~~~~~~~~~~~\nonumber\\
+\beta_{z}¥M_{z}¥^{4} +\beta_{yz}¥M_{z}¥^{2}¥M_{y}¥^{2}¥+\beta_{xz}¥M_{z}¥^{2}¥M_{x}¥^{2}-{\bf M}{\bf  H},
\label{F}
\end{eqnarray}
we bear in mind the orthorhombic anisotropy
but, unlike to Ref.10, the density of gradient energy will be  taken in exchange approximation
\begin{equation}
F_\nabla=\gamma_{ij}\frac{\partial {\bf M}}{\partial x_i}\frac{\partial {\bf M}}{\partial x_j}.
\label{nabla}
\end{equation}
Here, the $x, y, z$ are directions of the spin axes pinned to $a, b, c$
crystallographic directions correspondingly, 
\begin{equation}
\alpha_{z}=\alpha_{z0}¥(T-T_{c0}), 
\end{equation}
$\alpha_x>0$, $\alpha_y>0$ and matrix $\gamma_{ij}$ 
is 
\begin{equation}
\gamma_{ij} = \left(\begin{array}{ccc} \gamma_{xx} & 0 & 0\\
0 & \gamma_{yy} & 0 \\
0 & 0 & \gamma_{zz}
\end{array} \right).
\end{equation}

Let us take the magnetic field as the sum of the constant part and the coordinate dependent small addition 
\begin{equation}
{\bf H}({\bf r})=(H_x+\delta H_x({\bf r}))\hat x+(H_y+\delta H_y({\bf r}))\hat y+(H_z+\delta H_z({\bf r}))\hat z.
\end{equation}
By variation of functional (\ref{FE}) in respect to the components of magnetization we arrive to equations
\begin{eqnarray}
2\alpha_xM_x+2\beta_{xz}M^2_zM_x-\gamma_{ij}\frac{\partial^2 M_x}{\partial x_i\partial x_j}=H_x+\delta H_x,\\
\label{v}
2\alpha_yM_y+2\beta_{yz}M^2_zM_y-\gamma_{ij}\frac{\partial^2 M_y}{\partial x_i\partial x_j}=H_y+\delta H_y,
\label{va}\\
2\alpha_zM_z+4\beta_zM_z^3+2\beta_{xz}M_zM_x^2+2\beta_{yz}M_zM_y^2~~~~\nonumber\\-\gamma_{ij}\frac{\partial^2 M_z}{\partial x_i\partial x_j}=H_z+\delta H_z
\label{var}
\end{eqnarray}

In constant magnetic field ${\bf H}=H_y\hat y+H_z\hat z$
the equilibrium magnetization projections are determined by equations:
\begin{eqnarray}
M_y=\frac{H_y}{2(\alpha_y+\beta_{yz}M_z^2)},~~~~~~~~~~~~~~~~~~~~~~~~~~~~\\
M_z^2=\frac{1}{2\beta_z}\left (-\alpha_z-\frac{\beta_{yz}H_y^2}{4(\alpha_y+\beta_{yz}M^2_z)^2}\right )+\frac{H_z}{4\beta_zM_z}.
\end{eqnarray}
We see that under a magnetic field perpendicular to direction of spontaneous magnetization  the Curie temperature decreases as
\begin{equation}
T_c=T_c(H_y)=T_{c0}-\frac{\beta_{xz}H_y^2}{4\alpha_y^2\alpha_{z0}}.
\end{equation}
Thus, near the Curie temperature the $z$-component of magnetization is determined by the equation
\begin{equation}
M_z^2\cong\frac{\alpha_{0z}}{2\beta_z}(T_c(H_y)-T)+\frac{H_z}{4\beta_zM_z}.
\label{M}
\end{equation}
The first and the last term in this equation correspond to the spontaneous and the induced part of magnetization along $z$ direction. 
It should be noted that the Curie temperature is the critical temperature of second order phase transition in ferromagnetic state only at $H_z=0$. Otherwise it is a "temperature" of crossover from paramagnetic to ferromagnetic state under external field parallel to spontaneous magnetization.

One can obtain  similar formulas for the constant magnetic field oriented in $(x,z)$ plane ${\bf H}=H_x\hat x+H_z\hat z$.
This case 
\begin{eqnarray}
M_x=\frac{H_x}{2(\alpha_x+\beta_{xz}M_z^2)},~~~~\\
M_z^2\cong\frac{\alpha_{0z}}{2\beta_z}(T_c(H_x)-T)+\frac{H_z}{4\beta_zM_z},
\label{Mz}
\end{eqnarray}
\begin{equation}
T_c=T_c(H_x)=T_{c0}-\frac{\beta_{xz}H_x^2}{4\alpha_x^2\alpha_{z0}}.
\end{equation}

Experimentally  the Curie temperature is suppressed by magnetic field parallel to $b$ crystallographic axis \cite{Aoki2009,Hardy11}. In UCoGe the noticeable decrease  is started at fields about 5 Tesla such that at fields about 15 Tesla the Curie temperature falls to zero.\cite{Aoki2009}
On the contrary, for field in $a$- crystallographic direction a changes of $T_c$ till the field about 15 Tesla have not been revealed as it could be expected for the magnetically hardest axis. One can speculate, however, that the Curie temperature $T_c(H_x)$  begins a decrease at fields  higher than 20 Tesla.

Let us find first  the susceptibility  at finite field  in $(y,z)$ plane. 
Taking magnetization as the sum of the constant part and the coordinate dependent small addition 
\begin{equation}
{\bf M}({\bf r})=M_y\hat y+M_z\hat z+\delta M_y({\bf r})+\delta M_z({\bf r}),
\end{equation}
one can obtain from the equations (\ref{va}), (\ref{var}) the linear response of magnetization to the coordinate dependent part of magnetic field. The corresponding Fourier components of magnetization are
\begin{eqnarray}
\delta M_y({\bf k})=\chi_{yy}{\bf k})\delta H_y({\bf k})+\chi_{yz}({\bf k})\delta H_z({\bf k}),\\
\delta M_z({\bf k})=\chi_{zy}{\bf k})\delta H_y({\bf k})+\chi_{zz}({\bf k})\delta H_z({\bf k}),
\end{eqnarray}
where
\begin{eqnarray}
\chi_{yy}{\bf k})=\frac{c}{D},~~~~\chi_{zz}({\bf k})=\frac{a}{D},\\
\chi_{yz}({\bf k})=\chi_{zy}{\bf k})=-\frac{b}{D},
\end{eqnarray}
\begin{eqnarray}
a=2\alpha_y+2\beta_{yz}M_z^2+\gamma_{ij}k_ik_j=\frac{H_y}{M_y}+\gamma_{ij}k_ik_j,~~~~~~\\
b=4\beta_{yz}M_zM_y,~~~~~~~~~~~~~~~\\
c=8\beta_zM_z^2+\frac{H_z}{2M_z}+\gamma_{ij}k_ik_j,~~~~~~~~~~~~~~~~~\\
D=ac-b^2.~~~~~~~~~~~~~~~~~~~
\end{eqnarray}

Like in the paper \cite{Mineev2012} we shall use the following estimations for the constants in the GL free energy:
\begin{equation}
\beta_z\approx\frac{T_c}{2(m^2n)^2n},~~~~~~\gamma_{xx}\approx\gamma_{yy}\approx\gamma_{zz}\approx \frac{T_cr^2}{m^2n}.
\label{est1}
\end{equation}
Corresponding GL expression for the z-component of magnetization at $H_z=0$ is
\begin{equation}
M_z^2=(mn)^2\frac{T_c(H_y)-T}{T_c}.
\label{est2}
\end{equation}
Here $n$ is the concentration of uranium atoms with zero-temperature magnetic moment $m$, 
 $r$ is the distance between nearest neighbor uranium atoms.
 
The expressions for the susceptibility component are valid for small wave vectors $k<<k_F$.  We shall expect, however, that they are still qualitatively valid  at large wave vectors transfer $k\cong k_F$ determining pairing interaction. 
Then taking into account $k\cong k_F$ we come to estimation
\begin{equation}
\gamma_{ij} k_ik_j \approx \frac{T_c}{m^2n}.
\label{gamma}
\end{equation}

Let us assume 
\begin{equation}
ac>>b^2
\end{equation}
that is certainly true when $M_z$ is decreased together with the Curie temperature.  
Hence, one can work with the susceptibilities along  $c$-axis  and $b$ axis
in more simple form
\begin{eqnarray}
\chi_{zz}({\bf k})\cong\frac{1}{c}=
\frac{1}{8\beta_zM_z^2+\frac{H_z}{2M_z}+\gamma_{ij}k_ik_j},\\
\chi_{yy}{\bf k})\cong\frac{1}{a}= \frac{1}{\frac{H_y}{M_y}+\gamma_{ij}k_ik_j}=\frac{1}{\chi_y^{-1}+\gamma_{ij}k_ik_j}.
\end{eqnarray}
where the magnetization $M_z$ is determined by Eq.(\ref{M}) and $\chi_y=\frac{M_y}{H_y}$ is the constant part of susceptibility along $a$ direction. 

The same formula for $\chi_{zz}({\bf k})$ is valid  at finite field  in $(x,z)$ plane
but this case the magnetization $M_z$ is determined by Eq.(\ref{Mz}). For $x$-component of susceptibility
we have correspondingly
\begin{eqnarray}
\chi_{xx}{\bf k})\cong\frac{1}{a}= \frac{1}{\chi^{-1}_x+\gamma_{ij}k_ik_j}.
\end{eqnarray}

The odd part 
of $z$-component of susceptibility 
is found as 
\begin{eqnarray}
\chi^u_{zz}({\bf k},{\bf k}')=\frac{1}{2}[\chi_{zz}({\bf k}-{\bf k}')-\chi_{zz}({\bf k}+{\bf k}')]\nonumber\\
=\frac{2\gamma_{ij}k_ik_j^\prime}{
(8\beta_zM_z^2+\frac{H_z}{2M_z}+\gamma_{ij}(k_ik_j+k_i^{\prime}k_j^{\prime}))^2-(2\gamma_{ij}k_ik_j^{\prime})^2}.
\label{ch}
\end{eqnarray}
The pairing interaction is determined by this formula at ${\bf k}=k_F\hat k$. In that follows we shall  keep only angular dependence of interaction in numerator of Eq.(\ref{ch}) neglecting  by angular dependence of the orthorhombic symmetry terms in denominator $\gamma_{ij}(k_ik_j+k_i^{\prime}k_j^{\prime})\approx2\gamma k_F^2$
as well by all higher harmonics of interaction determined by the last term in denominator. Hence, we obtain
\begin{equation}
\chi^u_{zz}({\bf k},{\bf k}')\cong
\frac{2\gamma_{ij}k_i k_j^\prime}{(8\beta_zM_z^2+\frac{H_z}{2M_z}+2\gamma k_F^2)^2}.
\label{Chi}
\end{equation}

Found in similar manner the odd part 
of susceptibility $x$-component at finite field  in $(x,z)$ plane  is 
\begin{equation}
\chi^u_{xx}({\bf k},{\bf k}')\cong
\frac{2\gamma_{ij}k_i k_j^\prime}{(\chi_x^{-1}+2\gamma k_F^2)^2}.
\label{Chi3}
\end{equation}
The odd part 
of susceptibility $y$-component at finite field  in $(y,z)$ plane is 
\begin{equation}
\chi^u_{yy}({\bf k},{\bf k}')\cong
\frac{2\gamma_{ij}k_i k_j^\prime}{(\chi_y^{-1}+2\gamma k_F^2)^2}.
\label{Chi4}
\end{equation}

Thus, we have found all the susceptibility components determining the pairing interaction  besides 
$\chi^u_{xy}({\bf k},{\bf k}')$. The absence of off-diagonal susceptibility in orthorhombic crystal is direct consequence of exchange approximation we have used for the energy of magnetic inhomogeneity Eq.(\ref{nabla}). Out of  exchange approach the gradient energy in orthorhombic magnet contains also the invariants like
\begin{equation}
\gamma_{xy}\frac{\partial {M}_x}{\partial x}\frac{\partial { M}_y}{\partial y}.
\label{nabla1}
\end{equation}
This inevitably  leads to appearance of off-diagonal components of susceptibility 
\begin{equation}
\chi^u_{xy}({\bf k},{\bf k}')\cong-\frac{\gamma_{xy}(k_xk_y^\prime+k_x^\prime k_y)}{(\chi_x^{-1}+\gamma k_F^2)(\chi_y^{-1}+\gamma k_F^2)}.
\label{nabla1}
\end{equation}

\section{Field dependent pairing interaction}

For the pairing interaction Eq.(\ref{11}) we obtain
\begin{eqnarray}
V^{\uparrow\uparrow}({\bf k},{\bf k}')=V^{\downarrow\downarrow}({\bf k},{\bf k}')~~~~~~~~~~~~~~~~~~~~~~
\nonumber\\~~~~~~~=-
\frac{\mu_B^2I^2\gamma_{ij}k_ik_j^\prime}{2(\gamma k_F^2)^2(\frac{4\beta_zM_z^2}{\gamma k_F^2}+\frac{H_z}{4M_z \gamma k_F^2}+
1)^2}
\label{12}
\end{eqnarray}
The angular dependent pre-factor in this equation can be rewritten as follows
\begin{equation}
\frac{\mu_B^2I^2\gamma_{ij}k_ik_j^\prime}{ 2(\gamma k_F^2)^2}=\frac{T_c}{2n}\left (\frac{\mu_B}{m}\right )^2\tilde\gamma_{ij}\hat k_i\hat k_j^\prime,
\end{equation}
where $\tilde\gamma_{xx}=\gamma_{xx}/\gamma,\dots$ are numbers of the order of unity and $\hat k_i=k_i/k_F$ are components of the unit vector ${\bf k}$.
Thus, we obtain
\begin{eqnarray}
V^{\uparrow\uparrow}({\bf k},{\bf k}')=V^{\downarrow\downarrow}({\bf k},{\bf k}')=-V_1\tilde\gamma_{ij}\hat k_i\hat k_j^\prime\nonumber\\
=-V_1(\tilde\gamma_{xx}\hat k_x\hat k_x^\prime+\tilde\gamma_{yy}\hat k_y\hat k_y^\prime+\tilde\gamma_{zz}\hat k_z\hat k_z^\prime),
\label{13}
\end{eqnarray}
where
\begin{equation}
V_1=\frac{T_c}{2n}\left (\frac{\mu_B}{m}\right )^2\frac{1}{(\frac{4\beta_zM_z^2}{\gamma k_F^2}+\frac{H_z}{4M_z \gamma k_F^2}+
1)^2}.
\label{14}
\end{equation}
Similarly for the off-diagonal components we have
\begin{eqnarray}
V^{\uparrow\downarrow}({\bf k},{\bf k}')=-V_2\tilde\gamma_{ij}\hat k_i\hat k_j^\prime-iV_3
\tilde
\gamma_{xy}(k_xk_y^\prime+k_x^\prime k_y),
\label{15}\\
V^{\downarrow\uparrow}({\bf k},{\bf k}')=-V_2\tilde\gamma_{ij}\hat k_i\hat k_j^\prime+iV_3
\tilde\gamma_{xy}(k_xk_y^\prime+k_x^\prime k_y),
\label{151}
\end{eqnarray}
where
\begin{equation}
V_2=\frac{T_c}{2n}\left (\frac{\mu_B}{m}\right )^2\left (\frac{1}{(\frac{1}{2\chi_x\gamma k_F^2}+1
)^2}-\frac{1}{(\frac{1}{2\chi_y\gamma k_F^2}+1
)^2}\right ),
\label{16}
\end{equation}
\begin{equation}
V_3\cong\frac{T_c}{n}\left(\frac{\mu_B}{m}\right )^2\frac{1}{(\frac{1}{\chi_x\gamma k_F^2}+1
)(\frac{1}{\chi_y\gamma k_F^2}+1)},
\label{17}
\end{equation}
and $\tilde\gamma_{xy}=\gamma_{xy}/\gamma$.
 Due to relativistic smallness of coefficient $\tilde \gamma_{xy}$ in comparison with other coefficients $\tilde\gamma_{xx},\dots$ the imaginary parts in Eqs. (\ref{15}), (\ref{151}) are always much smaller than its real parts.  

One can see that the off-diagonal pairing amplitudes $V^{\uparrow\downarrow}$ and $V^{\downarrow\uparrow}$ are field independent, whereas the diagonal amplitudes $V^{\uparrow\uparrow}$ and $V^{\downarrow\downarrow}$ depend of magnetic field. 
Making use the estimations given by Eqs. (\ref{est1}), (\ref{est2}) for combination determining the denominator in Eq. (\ref{14}) at $H_z=0$ we have
\begin{equation}
\frac{4\beta_zM^2_z}{\gamma k_F^2}=2\frac{T_c(H_i)-T}{T_c},~~~~~~i=x,y.
\label{comb}
\end{equation}
Thus, at large enough magnetic field directed along either $a$ or $b $ axis this combination is small. 

It is difficult to find explicitly the  combinations $\frac{1}{2\chi_{x,y}\gamma k_F^2}$ determining the denominators in Eq. (\ref{16}).
In assumption that it is essentially larger than unity
 one can expect that 
 \begin{equation}
 V^{\uparrow\uparrow}>>V^{\uparrow\downarrow},
\label{neq}
 \end{equation} 
 otherwise these amplitudes are close in magnitude.

The explicit form of the pairing interaction is fixed and we can look now what kind of superconducting state it gives rise.

\section{Structure of superconducting state}

Even in the absence  of external field in ferromagnetic superconductors there is an internal field acting on the electron charges. Due to this reason the superconducting state is always inhomogeneous. So, working with Ginzburg-Landau-Gor'kov equations we should keep the gradient terms.

Performing the Taylor expansion of Eqs.(\ref{up}), (\ref{down}) in powers of ${\bf q}$ up to the second order and  and then transforming them  to the coordinate representation, that means simple substitution
\begin{equation}
{\bf q}\to {\bf D}=-i\nabla_{\bf r}+2e{\bf A}({\bf r}),
\end{equation}
we obtain  equations 
\begin{widetext}
\begin{eqnarray}
&\Delta^{\uparrow}({\bf k},{\bf r})=~~~~~~~~~~~~~~~~~~~~~~~~~~~~~~~~~~~~~~~~~~~~~~~~~~~~~~~~~~~~~~~~~~~~~~~~~~~~~~~~~~~~~~~~~~~~~~~\nonumber
\\
&T\sum_n\int\frac{d^3{\bf k}^\prime}{(2\pi)^3} V_1\tilde\gamma_{ij}\hat k_i\hat k_j^\prime\left(G^\uparrow({\bf k}',\omega_n)
G^\uparrow(-{\bf k}',-\omega_n)+\frac{1}{2}G^\uparrow({\bf k}',\omega_n)\frac{\partial^2G^\uparrow(-{\bf k}',-\omega_n)}{\partial k^\prime_l\partial k^\prime_m}D_lD_m
\right)\Delta^{\uparrow}({\bf k}^\prime,{\bf r})~~~~\nonumber\\
&+T\sum_{n}\int\frac{d^3{\bf k}^\prime}{(2\pi)^3}(V_2\tilde\gamma_{ij}\hat k_i\hat k_j^\prime+iV_3
\tilde
\gamma_{xy}(k_xk_y^\prime+k_x^\prime k_y))
\left(G^\downarrow({\bf k}',\omega_n)
G^\downarrow(-{\bf k}',-\omega_n)+\frac{1}{2}G^\downarrow({\bf k}',\omega_n)\frac{\partial^2G^\downarrow(-{\bf k}',-\omega_n)}{\partial k^\prime_l\partial k^\prime_m}D_lD_m
\right)\Delta^{\downarrow}({\bf k}^\prime,{\bf r}),~~\\
\label{up1}
&\Delta^{\downarrow}({\bf k},{\bf r})=~~~~~~~~~~~~~~~~~~~~~~~~~~~~~~~~~~~~~~~~~~~~~~~~~~~~~~~~~~~~~~~~~~~~~~~~~~~~~~~~~~~~~~~~~~~~~\nonumber\\
&T\sum_{n}\int\frac{d^3{\bf k}^\prime}{(2\pi)^3}(V_2\tilde\gamma_{ij}\hat k_i\hat k_j^\prime-iV_3
\tilde
\gamma_{xy}(k_xk_y^\prime+k_x^\prime k_y))
\left(G^\uparrow({\bf k}',\omega_n)
G^\uparrow(-{\bf k}',-\omega_n)+\frac{1}{2}G^\uparrow({\bf k}',\omega_n)\frac{\partial^2G^\uparrow(-{\bf k}',-\omega_n)}{\partial k^\prime_l\partial k^\prime_m}D_lD_m
\right)\Delta^{\uparrow}({\bf k}^\prime,{\bf r})\nonumber\\
&+T\sum_{n}\int\frac{d^3{\bf k}^\prime}{(2\pi)^3} V_1\tilde\gamma_{ij}\hat k_i\hat k_j^\prime\left(G^\downarrow({\bf k}',\omega_n)
G^\downarrow(-{\bf k}',-\omega_n)+\frac{1}{2}G^\downarrow({\bf k}',\omega_n)\frac{\partial^2G^\downarrow(-{\bf k}',-\omega_n)}{\partial k^\prime_l\partial k^\prime_m}D_lD_m
\right)\Delta^{\downarrow}({\bf k}^\prime,{\bf r}).
\label{down1}
\end{eqnarray}
\end{widetext}

Taking into account the angular structure of pairing interaction in orthorhombic ferromagnet given by equations (\ref{13}), (\ref{15}), (\ref{151}) we can choose the superconducting order parameter as the following linear combinations of 
momentum  direction projections on the coordinate axis
\begin{eqnarray}
\Delta^\uparrow({\bf k},{\bf r})=\hat k_x\eta_x^\uparrow({\bf r})+i\hat k_y\eta_y^\uparrow({\bf r})+\hat k_z\eta_z^\uparrow({\bf r}),\\
\Delta^\downarrow({\bf k},{\bf r})=\hat k_x\eta_x^\downarrow({\bf r})+i\hat k_y\eta_y^\downarrow({\bf r})+\hat k_z\eta_z^{\uparrow}({\bf r}).
\end{eqnarray}
Substituting these expressions to Eqs. (\ref{up1}) and (\ref{down1}) we come to the system of differential equations, that is convenient to present in the matrix form 
\begin{equation}
\eta_\alpha({\bf r})=A_{\alpha\beta}\eta_\beta({\bf r})
\end{equation}
for the components of vector
\begin{equation}
\eta_\alpha({\bf r})=(\eta_x^\uparrow({\bf r}),\eta_x^\downarrow({\bf r}),\eta_y^\uparrow({\bf r}),\eta_y^\downarrow({\bf r}),\eta_z^\uparrow({\bf r}),\eta_z^{\uparrow}({\bf r}))
\end{equation}
and the matrix operator
\begin{widetext}
\begin{eqnarray}
A_{\alpha\beta} =~~~~~~~~~~~~~~~~~~~~~~~~~~~~~~~~~~~~~~~~~~~~~~~~~~~~~~~~~~~~~~~~~~~~~~~~~~~~\nonumber\\
~~~~~~~~~~~~~~~~~~~~~~~~~~~~~~~~~~~~~~~~~~~~~~~~~~~~~~~~~~~~~~~~~~~~\nonumber\\
 \left(\begin{array}{cccccc}
 \medskip
g^\uparrow_{1x}\lambda+L_{1x}^\uparrow &
g_{2x}^\downarrow\lambda+L_{2x}^\downarrow+iL_{3yx}^\downarrow & 
iL_{1xy}^\uparrow & 
-g_{3xy}^\downarrow\lambda+iL_{2xy}^\downarrow-L_{3y}^\downarrow&
L_{1xz}^\uparrow &
L_{2xz}^\downarrow+iL_{3yz}^\downarrow\\
\medskip
g_{2x}^\uparrow\lambda+L_{2x}^\uparrow-iL_{3yx}^\uparrow & 
g^\downarrow_{1x}\lambda+L_{1x}^\downarrow  & 
g_{3xy}^\uparrow\lambda+iL_{2xy}^\uparrow +L_{3y}^\uparrow& 
iL_{1xy}^\downarrow& 
L_{2zx}^\uparrow-iL_{3yz}^\uparrow & 
L_{1xz}^\downarrow \\
\medskip
-iL_{1yx}^\uparrow & 
g_{3yx}^\downarrow\lambda-iL_{2yx}^\downarrow+L_{3x}^\downarrow &  
g_{1y}^\uparrow \lambda+L_{1y}^\uparrow& 
g_{2y}^\downarrow\lambda+L_{2y}^\downarrow+iL_{3xy}^\downarrow & 
 -iL_{1yz}^\uparrow& 
-iL_{2yz}^\downarrow+L_{3xz}^\downarrow \\
\medskip
-g_{3yx}^\uparrow\lambda-iL_{2yx}^\uparrow-L_{3x}^\uparrow& 
-iL_{1yx}^\downarrow & 
g_{2y}^\uparrow\lambda+L_{2y}^\uparrow-iL_{3xy}^\uparrow & 
g_{1y}^{\downarrow}\lambda+L_{1y}^\downarrow & 
-iL_{2yz}^\uparrow+L_{3xz}^\uparrow &
 -iL_{1yz}^\downarrow\\
 \medskip
 L_{1zx}^\uparrow &
 L_{2zx}^\downarrow & 
 iL_{1zy}^\uparrow& 
iL_{2zy}^\downarrow & 
 g_{1z}^\uparrow\lambda+L_{1z}^\uparrow & 
 g_{2z}^\downarrow\lambda+L_{2z}^\downarrow  \\
 \medskip
 L_{2zx}^\uparrow &
 L_{1zx}^\downarrow & 
 iL_{2zy}^\uparrow& 
iL_{1zy}^\downarrow & 
 g_{2z}^\uparrow\lambda+L_{2z}^\uparrow & 
 g_{1z}^\downarrow\lambda+L_{1z}^\downarrow 
 \end{array} \right).
\end{eqnarray}
\end{widetext}

Here, 
\begin{equation}
g_{1x}^\uparrow=V_1\langle\tilde\gamma_{xx}\hat k_x^2N_0^\uparrow(\hat {\bf k}) \rangle
\end{equation}
is one of  the constants of pairing interaction, the angular brackets mean the averaging over the Fermi surface, $N_0^\uparrow(\hat {\bf k})$ is the angular dependent density of electronic states at the Fermi surface of the band $\uparrow$. Correspondingly
\begin{equation}
g_{2x}^{\downarrow}=V_2\langle\tilde\gamma_{xx}\hat k_x^2N_0^\downarrow(\hat {\bf k}) \rangle,~~~~~~
g_{3xy}^{\downarrow}=V_3\langle\tilde\gamma_{xy}\hat k_y^2N_0^\downarrow(\hat {\bf k}) \rangle.
\end{equation}
All the other constants of pairing interaction  are obtained by obvious substitutions $x\leftrightarrow y$ and $\uparrow \leftrightarrow\downarrow$ or $x\rightarrow z$. 

The function 
\begin{equation}
\lambda(T)=2\pi T\sum_{n\geq 0}\frac{1}{\omega_n}=\ln\frac{\epsilon}{T},
\end{equation} where $\epsilon=\frac{2\gamma\varepsilon_0}{\pi}$, $\ln\gamma=0.577$ is the Euler constant, and $\varepsilon_0$  is an energy cutoff for pairing interaction. We assume here that it has the same value for both bands. 

The first type of differential operators is defined as follows
\begin{widetext}
\begin{equation}
L_{1x}^\uparrow=\frac{1}{2}V_1T\sum_n\int\frac{d^3{\bf k}}{(2\pi)^3} \tilde\gamma_{xx}\hat k_x^2
G^\uparrow({\bf k},\omega_n){\cal D},
\end{equation}
and $L_{2y}^\downarrow$ and the other operators with  same structure are obtained by obvious substitutions $(x\rightarrow y,z)$, $(1\rightarrow2)$ and $(\uparrow\rightarrow\downarrow)$, but  similar operator with index 3 is 
\begin{equation}
L_{3x}^\uparrow=\frac{1}{2}V_3T\sum_n\int\frac{d^3{\bf k}}{(2\pi)^3} \tilde\gamma_{xy}\hat k_y^2
G^\uparrow({\bf k},\omega_n){\cal D},
\end{equation}
here,
\begin{equation}
{\cal D}=
\frac{\partial^2G^\uparrow(-{\bf k},-\omega_n)}{\partial k_x^2 }D_x^2+\frac{\partial^2G^\uparrow(-{\bf k},-\omega_n)}{\partial k_y^2 }D_y^2+\frac{\partial^2G^\uparrow(-{\bf k},-\omega_n)}{\partial k_z^2 }D_z^2.
\end{equation}
The second type of operators is
\begin{equation}
L_{1xy}^\uparrow=\frac{1}{2}V_1T\sum_n\int\frac{d^3{\bf k}}{(2\pi)^3} \tilde\gamma_{xx}\hat k_x\hat k_y
G^\uparrow({\bf k},\omega_n)\frac{\partial^2G^\uparrow(-{\bf k},-\omega_n)}{\partial k_x\partial k_y }(D_xD_y+D_yD_x),
\end{equation}
and $L_{2yx}^\uparrow$ and the others operators of this type are obtained by obvious substitutions $(x\rightarrow y,z)$, $(1\rightarrow 2)$, $(\uparrow\rightarrow\downarrow)$, but similar  operators with index 3 are defined differently
\begin{equation}
L_{3xy}^\uparrow=\frac{1}{2}V_3T\sum_n\int\frac{d^3{\bf k}}{(2\pi)^3} \tilde\gamma_{xy}\hat k_x\hat k_y
G^\uparrow({\bf k},\omega_n)\frac{\partial^2G^\uparrow(-{\bf k},-\omega_n)}{\partial k_x\partial k_y }(D_xD_y+D_yD_x),
\end{equation}
\begin{equation}
L_{3yz}^\uparrow=\frac{1}{2}V_3T\sum_n\int\frac{d^3{\bf k}}{(2\pi)^3} \tilde\gamma_{xy}\hat k_y\hat k_z
G^\uparrow({\bf k},\omega_n)\frac{\partial^2G^\uparrow(-{\bf k},-\omega_n)}{\partial k_y\partial k_z }(D_yD_z+D_zD_y).
\end{equation}
\end{widetext}

\subsection{Field along the spontaneous magnetization}

For  the arbitrary direction of external magnetic field we have 6 coupled equations for the 6 order parameter components. The situation is   simplified   
in absence of external field, or when an external field is directed along the axis of spontaneous magnetization $\hat z$.
Then the long derivatives are
\begin{equation}
D_x=-i\frac{\partial}{\partial x},~~D_z=-i\frac{\partial}{\partial z},~~
D_y=-i\frac{\partial}{\partial y}+\frac{2e}{c}(H+H_{int})x.
\end{equation}
Here we have  introduced internal electromagnetic field corresponding spontaneous magnetization $H_{int}\approx mn$ and ignore  the difference between the external field and magnetic field induced inside the media by the external field.

Now  the order parameter components are $z$-coordinate independent. Putting all the $z$ derivative $D_z=0$ we come to the matrix $A_{\alpha\beta}$ in the form
\begin{widetext}
\begin{eqnarray}
 \left(\begin{array}{cccccc}
 \medskip
g^\uparrow_{1x}\lambda+L_{1x}^\uparrow &
g_{2x}^\downarrow\lambda+L_{2x}^\downarrow+iL_{3yx}^\downarrow & 
iL_{1xy}^\uparrow & 
-g_{3xy}^\downarrow\lambda+iL_{2xy}^\downarrow-L_{3y}^\downarrow&
0 &
0\\
\medskip
g_{2x}^\uparrow\lambda+L_{2x}^\uparrow-iL_{3yx}^\uparrow & 
g^\downarrow_{1x}\lambda+L_{1x}^\downarrow  & 
g_{3xy}^\uparrow\lambda+iL_{2xy}^\uparrow +L_{3y}^\uparrow& 
iL_{1xy}^\downarrow& 
0& 
0\\
\medskip
-iL_{1yx}^\uparrow & 
g_{3yx}^\downarrow\lambda-iL_{2yx}^\downarrow+L_{3x}^\downarrow &  
g_{1y}^\uparrow \lambda+L_{1y}^\uparrow& 
g_{2y}^\downarrow\lambda+L_{2y}^\downarrow+iL_{3xy}^\downarrow & 
 0& 
0 \\
\medskip
-g_{3yx}^\uparrow\lambda-iL_{2yx}^\uparrow-L_{3x}^\uparrow& 
-iL_{1yx}^\downarrow & 
g_{2y}^\uparrow\lambda+L_{2y}^\uparrow-iL_{3xy}^\uparrow & 
g_{1y}^{\downarrow}\lambda+L_{1y}^\downarrow & 
0&
 0\\
 \medskip
 0&
 0 & 
 0& 
0& 
 g_{1z}^\uparrow\lambda+L_{1z}^\uparrow & 
 g_{2z}^\downarrow\lambda+L_{2z}^\downarrow  \\
 \medskip
 0&
 0& 
 0& 
0& 
 g_{2z}^\uparrow\lambda+L_{2z}^\uparrow & 
 g_{1z}^\downarrow\lambda+L_{1z}^\downarrow 
 \end{array} \right).
 \label{syst}
\end{eqnarray}
\end{widetext}
We see that equations for  
\begin{eqnarray}
\Delta^\uparrow({\bf k},{\bf r})=\hat k_x\eta_x^\uparrow({\bf r})+i\hat k_y\eta_y^\uparrow({\bf r})
,\\
\Delta^\downarrow({\bf k},{\bf r})=\hat k_x\eta_x^\downarrow({\bf r})+i\hat k_y\eta_y^\downarrow({\bf r})
\end{eqnarray}
parts of the order paramer are decoupled from the equations for the
components
\begin{eqnarray}
\Delta^\uparrow({\bf k},{\bf r})=
\hat k_z\eta_z^\uparrow({\bf r}),\\
\Delta^\downarrow({\bf k},{\bf r})=
\hat k_z\eta_z^{\uparrow}({\bf r}),
\end{eqnarray}
as it should be because the terms  $\hat k_x\eta_x^{\uparrow,\downarrow}+i\hat k_y\eta_y^{\uparrow,\downarrow}$ correspond to the order parameter transforming according to $A_1$ co-representation of the point symmetry group of orthorhombic ferromagnet, whereas the terms $\hat k_z\eta_z^{\uparrow,\downarrow}$ belong to $B_1$ co-representation.\cite{Cham,Min04}

The critical temperatures of transition to superconducting state  $A_1$ or $B_1$ are defined by  the maximum of eigenvalue  of separate systems  differential equations for corresponding components of the order parameter.

Using the relativistic smallness of $\tilde\gamma_{xy}$ one can neglect in matrix Eq.(\ref{syst}) by all the terms with index 3. This case
the equations for the $x$ components of the order parameter are coupled with the equations for the $y$ components only due to the gradient terms like $L_{1xy}^\uparrow$ etc.  In absence of external filed  the gradient terms can be treated as small terms due to the smallness of internal electromagnetic field. Then the equations for $x$, $y$ and $z$ components of the order parameter are transformed into algebraic equations decoupled each other:
\begin{eqnarray}
\eta_x^\uparrow=(g_{1x}^\uparrow\eta_x^\uparrow+g_{2x}^{\downarrow}\eta_x^\downarrow)\Lambda,
\nonumber\\
\eta_x^\downarrow=(g_{2x}^{\uparrow}\eta_x^\uparrow+g_{1x}^\downarrow\eta_x^\downarrow
)\Lambda,
\label{Ax}
\end{eqnarray}
\begin{eqnarray}
\eta_y^\uparrow=(g_{1y}^\uparrow\eta_y^\uparrow+g_{2y}^{\downarrow}\eta_y^\downarrow)\Lambda,
\nonumber\\
\eta_y^\downarrow=(g_{2y}^{\uparrow}\eta_y^\uparrow+g_{1y}^\downarrow\eta_y^\downarrow
)\Lambda,
\label{Ay}
\end{eqnarray}
and
\begin{eqnarray}
\eta_z^\uparrow=(g_{1z}^\uparrow\eta_z^\uparrow+g_{2z}^{\downarrow}\eta_z^\downarrow)\Lambda,
\nonumber\\
\eta_z^\downarrow=(g_{2z}^{\uparrow}\eta_z^\uparrow+g_{1z}^\downarrow\eta_z^\downarrow
)\Lambda.
\label{Bz}
\end{eqnarray}
Thus, in the exchange approximation for the energy of magnetic inhomogeneity we have three different superconducting states $(\hat k_x\eta_x^\uparrow,~\hat k_x\eta_x^\downarrow),$  $(\hat k_y\eta_y^\uparrow,~\hat k_y\eta_y^\downarrow)$ and $(\hat k_z\eta_z^\uparrow,~\hat k_z\eta_z^\downarrow)$ with different critical temperatures defined by the determinants of Eqs.  (\ref{Ax}), (\ref{Ay}) and (\ref{Bz}).

\section{Field dependent critical temperature}

Let us discuss the situation in neglect the orbital effects that is 
as if the magnetic field acts only on the electron spins.

Assuming that the largest critical temperature corresponds to the $(\hat k_x\eta_x^\uparrow,~\hat k_x\eta_x^\downarrow)$ superconducting state the zero of determinant of the system (\ref{Ax}) yields the BSC-type formula
\begin{equation}
T=\epsilon~exp\left (-\frac{1}{g}  \right ),
\label{21}
\end{equation}
where the constant of interaction
\begin{equation}
g=\frac{g_{1x}^\uparrow+g_{1x}^\downarrow}{2}+\sqrt{\frac{(g_{1x}^\uparrow-g_{1x}^\downarrow)^2}{4}+g_{2x}^{\uparrow}g_{2x}^{\downarrow}}
\end{equation}
is the function of temperature and magnetic field.
The formula (\ref{21}) is, in fact the equation for the determination of the critical temperature of the transition to the superconducting state $T_{sc}=T_{sc}({\bf H})$. 

As we have seen in the Chapter IV the constants $g_{1x}^\uparrow$ and $g_{1x}^\downarrow$ are field dependent and the constants $g_{2x}^{\uparrow}$ and $g_{2x}^{\downarrow}$ are not. The Curie temperature in URhGe at zero field is about 10 K, and it is practically field independent in magnetic fields smaller than the upper critical field both along a and $b$ directions \cite{Hardy} whereas the superconductivity is developed at temperatures below $T_{sc}\approx 0.25$ K. It means that the combination in equation (\ref{comb}) $\frac{4\beta_zM_z^2}{\gamma k_F^2}\approx 2$ and all the pairing amplitudes are  proved to be field independent. Thus the critical 
temperature of superconducting phase transition is independent of magnetic field.

Another situation is realized in UCoGe in field parallel to $b$-axis.  It is impossible  to say something definite about the absolute value of the pairing constants. However, one can easily compare their relative values  in different field regions.
 It is simplest to consider this in the case of validity of inequality (\ref{neq}) that corresponds to single band superconductivity.
Then  the critical temperature  in field parallel to $b$-axis is determined by 
\begin{equation}
T=\epsilon~exp\left (-\frac{1}{g_{1x}^\uparrow}  \right ),
\label{22}
\end{equation}
where according to Eqs. (\ref{14}) and (\ref{comb}) 
\begin{equation}
g_{1x}^\uparrow\propto\frac{1}{(2\frac{T_c(H_y)-T}{T_c}+1)^2}.
\end{equation}
In zero magnetic field $T_c=3.5$ K and $T_{sc}=0.55$ K,
hence, in small enough magnetic fields, so long  the Curie temperature is field independent,
$\frac{T_c(H_y)-T}{T_c}\approx 1$ and 
the critical  temperature 
of superconducting transition is field independent as it is in URhGe.
However, for field $H_y> 5$ Tesla the Curie temperature begins sharp decreasing, the combination 
$\frac{T_c(H_y)-T}{T_c}$ starts to be less and less than 1. This leads to the sharp increase of the pairing constant $g_x^\uparrow$ and the effective superconducting critical  temperature grows exponentially, that corresponds to the robustness of superconducting state in field interval $5<H_y<10$ Tesla.\cite{Aoki2009}

Similar mechanism also stimulates the  arising the reentrant superconducting state in URhGe.   
Indeed, the Curie temperature in fields along $b$-axis about 10 Tesla  is significantly smaller than it is in zero field.\cite{Hardy11} This leads to drastic increase of pairing interaction
and reappearance of the superconducting state reported in Ref.13. One must stress that the reentrant superconducting state appears already at field about 8 Tesla that is 
in significantly lower  field $H_b$ than the field of the first order transition equal to 12 Tesla reported in papers \cite{Levy2007,Levy2009}.

The magnetic anisotropy in $a$ direction is much harder and a decrease of the Curie temperature in fields $H_x$ up to 
10-12 Tesla has not yet reported. However,  at low temperatures when the upper critical field along $a$ direction is
larger than 10 Tesla  the upper critical field in UCoGe reveals the upward  curvature. This in our opinion corresponds to the beginning of suppression of the Curie temperature and corresponding increase of the pairing interaction like it is in smaller fields directed along $b$ axis.

The above discussion is applicable to the situations with the external field parallel either $b$ or $a$ axis. In presence also the field $H_z$  along the direction of spontaneous magnetization one must work with 
pairing constant $g_{1x}^\uparrow$ proportional to pairing potential $V_1$ given by  general formula (\ref{14}) where $M_z$ is determined either by Eq. (\ref{M}) or by Eq. (\ref{Mz}). One can see that the field along $c$ axis increases the magnetic moment and definitely decreases the pairing amplitude $g_{1x}^\uparrow$.  This is probably the source of non GL sharp anisotropy of the upper critical field in UCoGe observed \cite{Aoki2009} at very low temperature T=0.85 K at quite small field  inclinations from $a$-axis direction toward $c$ axis.

\section{Conclusion}

We have derived the relationship between the magnetic  and superconducting properties in uranium ferromagnetic superconductors.  It was shown that  the long time ago established  general form of the superconducting order parameter dictated by symmetry \cite{Cham,Min04} directly follows from the microscopic model where the interaction between the electrons is supported by the magnetization fluctuations in ferromagnetic metal with orthorhombic symmetry. 

The suppression of the Curie temperature by the magnetic field in $b$-direction   results 
 in the strong robustness of superconducting state in UCoGe at fields exceeding $\approx 5$ Tesla. The same mechanism is probably responsible for the low temperature upturn of the upper critical field  parallel to $a$-axis in the same material reported in Ref.11.

The developed theory allows to explain several basic properties of uranium ferromagnetic superconductors like the order parameter structure, presence of nodes in quasiparticle spectrum and  temperature dependence of upper critical field. Certainly, there are some properties of these materials like field dependent effective mass \cite{Aoki14}
and strong anisotropy of NMR and NQR attenuation \cite{T.Hattory} 
which could find an explanation in more elaborate strong coupling theory. However, even our simplified treatment demonstrate that due to multi-band and multi-component character of superconductivity to reach full quantitative description of superconducting state in these materials is by no means a simple problem.

\section*{Acknowledgements}

I am grateful to K. Ishida  for the kind possibility to be aware of his recent NMR results.


\begin{thebibliography}{20}
\bibitem{footnote} The case  of UCoGe at pressure about 1.5 GP when the Curie temperature is almost equal to the temperature of superconducting transition demands the particular treatment.
\end{thebibliography}
\end{document}